\documentclass[sigconf]{acmart}

\usepackage{multirow}
\usepackage{subcaption}
\usepackage{xcolor}
\usepackage{soul}

\definecolor{mycolor}{HTML}{4C98FF} %

\DeclareRobustCommand{\nicenumber}[1]{\sethlcolor{mycolor}\textcolor{white}{\textbf{\hl{#1}}}}

\AtBeginDocument{%
  }

\copyrightyear{2026}
\acmYear{2026}
\setcopyright{cc}
\setcctype{by-nc-nd}
\acmConference[IUI '26]{31st International Conference on Intelligent User Interfaces}{March 23--26, 2026}{Paphos, Cyprus}
\acmBooktitle{31st International Conference on Intelligent User Interfaces (IUI '26), March 23--26, 2026, Paphos, Cyprus}
\acmPrice{}
\acmDOI{10.1145/3742413.3789132}
\acmISBN{979-8-4007-1984-4/2026/03}

\begin{document}

\title{Mapping the Design Space of User Experience for Computer Use Agents}

\author{Ruijia Cheng}
\affiliation{%
  \institution{Apple}
  \city{Seattle}
  \state{WA}
  \country{USA}}
\email{rcheng23@apple.com}

\author{Jenny T. Liang}
\authornote{Work completed while at Apple}
\affiliation{%
  \institution{Carnegie Mellon University}
  \city{Pittsburgh}
  \state{PA}
  \country{USA}
}
\email{jtliang@cs.cmu.edu}

\author{Eldon Schoop}
\affiliation{%
  \institution{Apple}
  \city{Seattle}
  \state{WA}
  \country{USA}}
\email{eldon@apple.com}

\author{Jeffrey Nichols}
\affiliation{%
  \institution{Apple}
  \city{Seattle}
  \state{WA}
  \country{USA}}
\email{jwnichols@apple.com}
\renewcommand{\shortauthors}{Cheng et al.}

\begin{abstract}

Large language model (LLM)-based computer use agents execute user commands by interacting with available UI elements, but little is known about how users want to interact with these agents or what design factors matter for their user experience (UX). We conducted a two-phase study to map the UX design space for computer use agents. In Phase 1, we reviewed existing systems to develop a taxonomy of UX considerations, then refined it through interviews with eight UX and AI practitioners. The resulting taxonomy included categories such as user prompts, explainability, user control, and users’ mental models, with corresponding subcategories and example design features. In Phase 2, we ran a Wizard-of-Oz study with 20 participants, where a researcher acted as a web-based computer use agent and probed user reactions during normal, error-prone and risky execution. We used the findings to validate the taxonomy from Phase 1 and deepen our understand of the design space by identifying the connections between design areas and divergence in user needs and scenarios. Our taxonomy and empirical insights provide a map for developers to consider different aspects of user experience in computer use agent design and to situate their designs within users' diverse needs and scenarios.

\end{abstract}

\ccsdesc[500]{Human-centered computing~Interactive systems and tools}

\keywords{computer use agents, user experience, formative research, Wizard-of-Oz}

\begin{teaserfigure}
  \includegraphics[width=\textwidth]{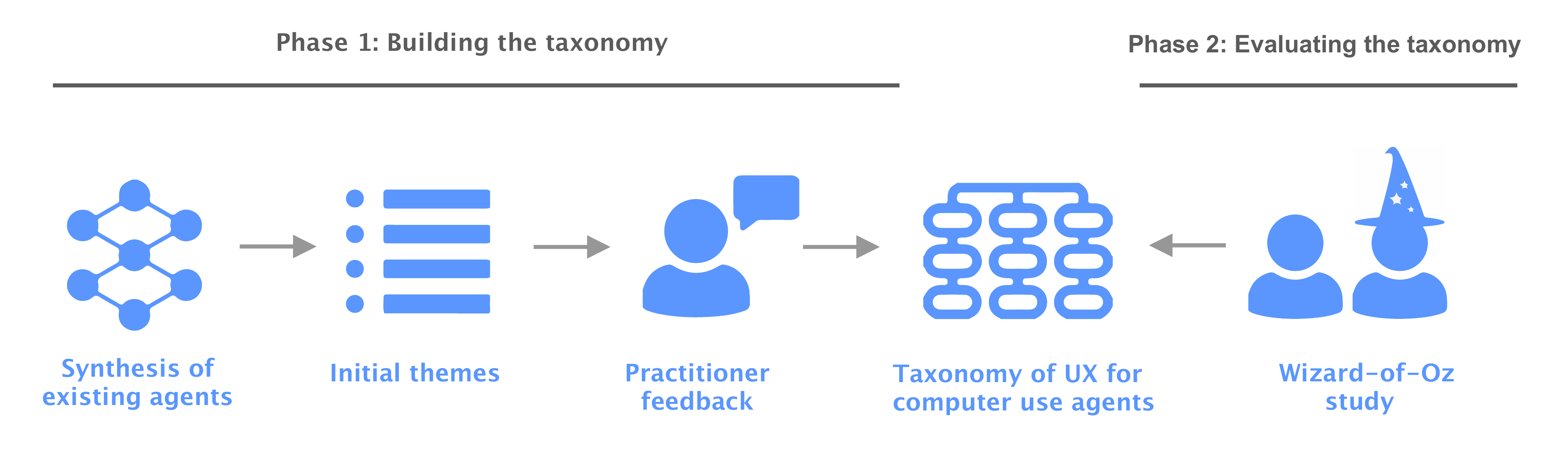}
  \caption{An overview of our two-phase study to map the design space of user experience for computer use agents. In Phase 1, we reviewed nine recently released computer use agents and synthesized initial themes in their design related to user experience. These themes were then refined through feedback and iteration with practitioners, resulting in a taxonomy of UX considerations for computer use agents. In Phase 2, we evaluated and deepened our understanding of the taxonomy
 through a Wizard-of-Oz study with 20 participants.}
 \Description{An overview diagram of our two-phase study to map the design space of user experience for computer use agents. Phase 1 is building the taxonomy, which consists of a series of sequential activities, including synthesis of existing agents, formation of initial themes of the taxonomy based on insights from existing agents, collection of feedback from practitioners on the initial themes. This leads to the taxonomy of UX for computer use agents. Phase 2 is evaluating the taxonomy, which consists of a Wizard-of-Oz study that validates and provides empirical insights about the taxonomy.}
  \label{fig:teaser}
\end{teaserfigure}

\maketitle

\section{Introduction}
Intelligent \textit{computer use agents}, also referred to as ``UI control agents,'' have emerged rapidly in the last year \cite{anthropic_computer_use_tool_2025, mozannar2025magenticuihumanintheloopagenticsystems,qin2025uitarspioneeringautomatedgui,openai_operator_2025}. Powered by Large Language Models (LLMs) or Multimodal Large Language Models (MLLMs), these agents execute user commands by interacting with available UI elements, such as by tapping buttons or scrolling pages \cite{tang2025survey}. While significant research has focused on developing agent models and evaluative benchmarks \cite{agrawal2025uinavbench, li2025ferretui2masteringuniversal, jiang2025ILuvUI, tang2025survey}, few have studied how human users might want to interact with and make use of these agents, and even fewer have investigated how user interaction with a computer use agent could be designed or what factors should be considered. To address this gap, we investigate: \emph{What is the design space of user experience (UX) for computer use agents?} 

In this paper, we map the design space in a two-phase formative study. In Phase 1, we reviewed existing systems to develop a taxonomy that categorizes different aspects related to UX. We analyzed 9 existing computer use agents that were announced in 2024 and 2025 and developed a set of categories describing the design space. We then conducted an interview study with a diverse group of eight practitioners in UX and AI, in which we presented our initial taxonomy and collected their feedback. This resulted in refinements to the taxonomy, including new wording to describe categories and modifications to the categories and their sub-categories. The final design space includes the following areas: user prompts, explainability of agent activities, user control, and the user's mental model of computer use agents. These areas are then sub-categorized, as we will describe later in the paper.

While Phase 1 focused on building the taxonomy from existing computer use agents and practitioner feedback, Phase 2 sought to further evaluate the taxonomy's coverage and examine which aspects participants valued across different contexts. To this end, we conducted a Wizard-of-Oz study with 20 participants using a browser-based computer use agent. A researcher acted as the web browser–based computer use agent and carried out common tasks based on user commands while probing for participants’ reactions and needs during execution. Scenarios involving errors and risks were also considered. In phase 2, we validated the coverage of the taxonomy and gained a deeper understanding of end users’ experiences and needs when interacting with computer use agents.

In summary, this paper contributes a taxonomy of UX considerations for computer use agents, developed through an analysis of existing systems, iterated with practitioner feedback, and validated through a Wizard-of-Oz study with end users. In addition, we contribute empirical insights into users’ experiences and expectations when interacting with computer use agents, enriching our understanding of the design space by identifying the connections between design areas and divergence in user needs and scenarios. 
We expect the taxonomy to be useful in helping a designer balance the factors and features that arise when designing a new computer use agent. 
We also invite future work to identify use cases of computer use agents that are not included in this taxonomy, and to contribute new dimensions that extend and refine it.

\section{Related Work}

\subsection{Computer use agents}
The idea of computer use agents---software systems that perform tasks through interacting with UIs---has deep roots in the HCI community. For several decades, researchers have explored interface agents that can autonomously change user interface states at a user's command \cite{lieberman1997autonomous}. Beyond autonomous operation, prior work has also investigated how to combine agent-based control with direct user manipulation, enabling collaboration between users and agents \cite{shneiderman1997direct, horvitz1999principle}. 

Early applications of the idea of computer use agents spanned a variety of domains. One notable example was CoScripter \cite{leshed2008coScripter}, a browser plug-in that enabled end users to record and replay actions in a web browser through human-readable scripts. %
Vegemite extended CoScripter with features specifically designed to support data analysis tasks \cite{lin2009vegemite}.
ScriptAgent leveraged programming by example to record procedures that could be generalized by the agent \cite{lieberman1998integrating}. \citet{kozierok1993learning} introduced an agent that applied reinforcement learning to adapt to a user’s personal scheduling preferences through observation and feedback. Similarly, \citet{lieberman1997autonomous} supported web browsing by conducting a concurrent breadth-first exploration based on the current navigation state.

While these early systems laid the foundations for autonomous UI control, %
recent advances in LLMs and MLLMs have enabled a new generation of computer use agents---autonomous systems capable of understanding, navigating, and interacting with digital interfaces \cite{tang2025survey}.
Examples of these systems include ILuvUI \cite{jiang2025ILuvUI} and Spotlight \cite{Li2022SpotlightMU}, which focus on single-screen UI tasks such as screen summarization and widget interaction. Ferret-UI \cite{ferretui-mobile} and its successor Ferret-UI 2 \cite{li2025ferretui2masteringuniversal} extend these capabilities with grounding and reasoning for UI interaction. Other efforts have targeted web environments, such as WebVoyager \cite{he-etal-2024-webvoyager}, OS-Copilot \cite{wu2024oscopilot}, CowPilot \cite{huq-etal-2025-cowpilot}, and OpenWebAgent \cite{iong-etal-2024-openwebagent}, with features ranging from autonomous browsing to user-supervised execution. AutoDroid \cite{wen2024AutoDroid} combines commonsense knowledge from LLMs with domain-specific knowledge of Android apps to enable general task automation, and AXNav \cite{taeb2024AXNav} applies LLM agents to natural language–based accessibility testing.

While much of the current research has focused on model design, benchmarks, and evaluation techniques \cite{agrawal2025uinavbench}, little attention has been paid to the user experience of interacting with computer use agents. Yet, prior work on pre-(M)LLM systems has shown that user experience is a critical dimension. \citet{shneiderman1997direct} argued that ``agents are not alternatives to direct manipulation,'' while \citet{horvitz1999principle} proposed that mixed-initiative systems must carefully balance agent autonomy with user control. Important considerations include how to provide users with sufficient visibility into the agent’s actions, when to allow users to control the agent, how to represent the agent within the interface, and what modalities the agent can accept \cite{SCHIAFFINO2004129, shneiderman1997direct, horvitz1999principle}.
With the emergence of LLM- and MLLM-powered agents, these challenges are amplified. Their greater autonomy and generality introduce new questions about trust, transparency, delegation, and control \cite{bansal2024challengeshumanagentcommunication}. 

Understanding the space of user experience in this new generation of computer use agents is essential for designing systems that are not only capable but also usable. Only recently, emerging LLM-based computer use agents start to explicitly investigate user experience. For example, CowPilot \cite{huq-etal-2025-cowpilot} introduces lightweight forms of user control by allowing users to pause and intervene before the agent executes web tasks. However, there exists no comprehensive mapping of the design space for user experience in computer use agents. Our work addresses this gap by specifically examining the user experience of computer use agents.

\subsection{The design space of user experience in human-AI collaboration}

Beyond computer use agents, there is a rich body of work exploring the user experience design space for human–AI collaboration more broadly. Much of this research has focused on developing design principles and guidelines. For example, \citet{amershi2019guidelines} provides one of the most comprehensive sets of guidelines for human–AI collaboration, emphasizing issues such as transparency, feedback, and user control. %
More recently, \citet{weisz2024design} synthesizes insights from multiple domains of human–AI interaction to propose six design principles tailored to generative AI systems, including to design with responsibility, for generative variability, mental models, co-creation, appropriate trust and reliance, and imperfection. We draw on these broader design dimensions in human–AI interaction when developing our taxonomy of user experience for computer use agents.

Another body of work relevant to our study comes from the user perspectives of general, not UI-based, agentic systems. For example, \citet{bansal2024challengeshumanagentcommunication} synthesizes challenges in human–agent communication, including how agents should help users verify their behavior, convey consistency, choose appropriate levels of detail, and decide which past interactions to consider when communicating, highlighting the importance of conveying both the agent’s activity to the user and the user’s intent to the agent.
\citet{feng2025levels} offers a complementary perspective on user roles representing different levels of user involvement when interacting with an agent, ranging from observer and approver to consultant that gives agents feedback, collaborator that co-plans and executes with agents, and operator who directs and makes all decisions. We also draw from emerging literature examining user involvement in LLM-driven agents. For instance, \citet{he2025plan-Then-Execute} explores ways to support user participation during both planning and execution, outlining intervention mechanisms such as splitting tasks, deleting steps, or confirming and rejecting actions. Similarly, \citet{feng2025cocoacoplanningcoexecutionai} offers co-planning mechanisms, where users can edit and re-prompt the agent’s plan, and co-execution strategies, where users and agents alternate in carrying out tasks. Finally, \citet{zhang2025from} provides the only known analysis of computer use agents from a user-centered perspective, categorizing potential risks, unintended consequences, and impacts that such agents may impose on users. 

The development of our taxonomy is inspired by these existing work that explores the design spaces of human-AI interaction. According to \citet{tang2025survey}, computer use agents can be decomposed into components of perception, exploration, planning, and interaction, and we believe each of which introduces considerations for user experience that are unique to the nature of computer use agents. Our taxonomy also took inspiration from \citet{zhang2025from}'s taxonomy of risks, impacts, and unintended consequences of computer use agents. We follow the iterative taxonomy development approaches that combine review of existing systems and human feedback (such as in \citet{amershi2019guidelines}, \citet{cheng2022mapping}, and \citet{zhang2025from}). 

\section{Phase 1: Building the taxonomy of UX for computer use agents}
\label{sec:taxonomy}
\subsection{Methods}
\subsubsection{Developing the initial taxonomy based on existing computer use agents}
We started the development of the taxonomy by reviewing 9 computer use agents that were announced to the public in 2024 to 2025 as shown in Table \ref{tab:ui_agents}. %
These agents represent the state of the art in UI control and operate across a range of platforms, including desktop, browser, and mobile. For each agent, the research team reviewed the demo videos if available, and installed and used the agent for 30 minutes for the ones that were available. We took field notes during the demo videos and usage sessions, and discussed as a team the significant themes that we observed. 

As a result of those discussions, we came up with an initial taxonomy containing a categorization of the design space for LLM-based computer use agents. The taxonomy contains themes and sub-themes of design considerations related to user experience, specifically: user prompts, explainability of agent activities, user control, end of task indicators, and task preview. For each of the themes, we also included examples of design features sourced from the set of computer use agents that we reviewed. We also included speculative design features that the research team imagined but did not see among the existing computer use agents.

\begin{table}[h]
\centering
\begin{tabular}{lp{6cm}}
\toprule
\textbf{Platform} & \textbf{Agents} \\
\hline
Desktop & Claude Computer Use Tool \cite{anthropic_computer_use_tool_2025}, Adept \cite{adept_ai_2025}, \newline
OpenAI Operator \cite{openai_operator_2025}, AIlice \cite{myshell_ailice_2025}, Magentic-UI \cite{mozannar2025magenticuihumanintheloopagenticsystems}, UI-TARS (Computer) \cite{qin2025uitarspioneeringautomatedgui} \\
\hline
Web Browser & Project Mariner \cite{deepmind_mariner_2025}, TaxyAI \cite{taxyai_browser_extension_2025}, 
\newline
AutoGLM (Browser) \cite{liu2024autoglmautonomousfoundationagents} \\
\hline
Mobile & UI-TARS (Mobile) \cite{qin2025uitarspioneeringautomatedgui}, AutoGLM (Mobile) \cite{liu2024autoglmautonomousfoundationagents} \\
\bottomrule
\end{tabular}
\caption{Computer use agents reviewed for the development of the taxonomy.}
\label{tab:ui_agents}
\end{table}

\subsubsection{Iterating the taxonomy}
We then presented our initial version of the taxonomy in an interview study with 8 practitioners who are designers, engineers, or researchers working in the domains of UX or AI at a large technology company. We recruited the participants through internal email lists, message channels, and snowball sampling. Our study protocol with participants were reviewed and approved by a research ethics committee for human subjects and legal counsel at our company. The goal of the study was to collect feedback on our initial taxonomy, specifically how each theme aligns with design considerations, whether there are any important omissions, and how it could be extended with sub-themes (e.g., for the theme about agent mistakes, what are the different types of agent mistakes?).  

Each of the interview study sessions lasted 45 to 60 minutes. During each session, a researcher first asked the participant about an example of LLM-based computer use agent which they had used or seen demos of. The goal was to ground their thinking around computer use agents and probe for UX-related considerations. Example questions included: 
What kind of tasks do you expect a user to achieve with this agent? 
What are some challenges that you anticipate a user to face when interacting with the agent? 

We then showed our initial categorization to participants via a slide deck. We explained each theme and design feature, supplemented with screenshots and video snippets from the agents that we reviewed. We asked for feedback with the specific goal of answering the following questions: 

\begin{itemize}
    \item Are there missing themes and sub-themes that are important to user experience in computer use agents? 
    \item How important are each of these themes and sub-themes to the design of computer use agents? 
    \item How can each of the design features potentially impact the user experience? 
\end{itemize}

We recorded the audio and video of each session and took field notes. We transcribed the recordings and conducted thematic analysis. Based on the feedback from participants, we iterated our original version of the taxonomy by re-organizing and renaming the themes, as well as adding new themes, into a final taxonomy.

\subsection{The taxonomy of UX for computer use agents}

Table \ref{tab:taxonomy} presents the complete taxonomy, which follows a three-level hierarchy including \textit{category}, the high-level areas of UX considerations in the design space; \textit{subcategory}, the specific aspects to consider in a particular area; and \textit{example features}, illustrative features supporting the UX consideration in a subcategory that are observed in existing agents or speculated by the research team.

\begin{table*}[]
\footnotesize
\begin{tabular}{|l|l|l|}
\hline
\textbf{Category} & \textbf{Subcategory} & \textbf{Example feature} \\ \hline
\multirow{17}{*}{User query} & \multirow{2}{*}{Levels of expression} & Speaking the gesture \\ \cline{3-3} 
 &  & Expressing the intent \\ \cline{2-3} 
 & \multirow{2}{*}{When the user enter the query} & Single query in the beginning \\ \cline{3-3} 
 &  & Conversational interaction \\ \cline{2-3} 
 & Modality of user input & Text, images,
voice, and others \\ \cline{2-3} 
 & \multirow{3}{*}{User profile} & Activity history in the app \\ \cline{3-3} 
 &  & Familiarity with the UI \\ \cline{3-3} 
 &  & Global preferences \\ \cline{2-3} 
 & \multirow{3}{*}{Ambiguity} & User mistakes \\ \cline{3-3} 
 &  & Missing parameters \\ \cline{3-3} 
 &  & Multiple references on the UI \\ \cline{2-3} 
 & \multirow{3}{*}{Contextual factors} & Device type \\ \cline{3-3} 
 &  & User goals \\ \cline{3-3} 
 &  & User mental state \\ \cline{2-3} 
 & \multirow{3}{*}{Safety} & Safety policy \\ \cline{3-3} 
 &  & Guardrails on whether the query is allowed \\ \hline
\multirow{14}{*}{Explainability of Agent Activities} & \multirow{2}{*}{Visibility of agent activities} & Where the action occurs on the UI \\ \cline{3-3} 
 &  & Retainability \\ \cline{2-3} 
 & \multirow{2}{*}{Description of agent action} & Description of the action \\ \cline{3-3} 
 &  & Whether and how other tools are called \\ \cline{2-3} 
 & \multirow{3}{*}{Transparency of agent reasoning} & Description of thoughts \\ \cline{3-3} 
 &  & Knowledge base consulted \\ \cline{3-3} 
 &  & Uncertainty of the model \\ \cline{2-3} 
 & \multirow{2}{*}{Preview of next steps} & Process behind decision \\ \cline{3-3} 
 &  & Showing next action \\ \cline{2-3} 
 & \multirow{2}{*}{Presentation of plan} & Overall plan \\ \cline{3-3} 
 &  & Replanning \\ \cline{2-3} 
 & \multirow{3}{*}{Communication of runtime status} & Confirmation on the execution of the last step \\ \cline{3-3} 
 &  & Success/failure at the end of the task \\ \cline{3-3} 
 &  & Notification of runtime status \\ \hline
\multirow{15}{*}{User control} & \multirow{4}{*}{User intervention during agent execution} & Stop/pause \\ \cline{3-3} 
 &  & User taking over control \\ \cline{3-3} 
 &  & User demonstration \\ \cline{3-3} 
 &  & Revert to previous steps \\ \cline{2-3} 
 & \multirow{3}{*}{High impact scenarios} & Indicators about the risk/impact \\ \cline{3-3} 
 &  & User permission to proceed \\ \cline{3-3} 
 &  & Levels of impact \\ \cline{2-3} 
 & \multirow{3}{*}{User intervention on plan} 
 & Review and edits on overall plan \\ \cline{3-3} 
 &  & User involvement in replanning \\ \cline{2-3}
 & \multirow{5}{*}{Agent error} & Type of agent errors \\ \cline{3-3} 
 &  & Error recovery \\ \cline{3-3} 
 &  & Error discoverability \\ \cline{3-3} 
 &  & Error communication \\ \hline
\multirow{13}{*}{User mental model \& expectations} & \multirow{4}{*}{Agent capability} & Type of tasks that the agent can perform \\ \cline{3-3} 
 &  & Supported type of user queries \\ \cline{3-3} 
 &  & Supported user controls \\ \cline{3-3} 
 &  & Supported user goals \\ \cline{2-3} 
 & \multirow{4}{*}{UI context} & UI constraints on agent \\ \cline{3-3} 
 &  & User's understanding of the UI \\ \cline{3-3} 
 &  & User and agent's interaction with the UI \\ \cline{2-3} 
 & \multirow{2}{*}{Scope of the agent} & External resources accessible by the agent \\ \cline{3-3} 
 &  & Scope of the UI that the agent has control over \\ \cline{2-3} 
 & \multirow{3}{*}{Risks} & Personal data and privacy \\ \cline{3-3} 
 &  & Type of potential risks brought by the agent \\ \cline{3-3} 
 &  & Mitigation mechanisms \\ \hline
\end{tabular}
\caption{The taxonomy that maps the areas of UX design considerations for computer use agents. }
\label{tab:taxonomy}
\end{table*}

The taxonomy covers four categories of UX considerations in the design space: \textit{User query}, \textit{Explainability of agent activities}, \textit{User control}, and \textit{Mental model}. Each category, subcategory, and the corresponding example features are explained below. 

\subsubsection{User Query} 
This category covers the design considerations involved in supporting users as they input commands to the agent. It includes the following subcategories:  

The design of the user experience for issuing commands to an agent should account for the varying \textbf{levels of expression} users may employ in their query. For instance, a user might directly speak the gesture by stating the action and the UI element (e.g., ``Click the back button''). Alternatively, the user might express their intent in a more semantic way (e.g., ``I want to go back to the last page'').  

It is also important to consider \textbf{when the user enters a query} during their interaction with the agent. In some cases, the user may provide a single query at the beginning and expect the agent to carry out the task autonomously. Alternatively, the user may anticipate a more conversational interaction, entering follow-up queries for corrections or confirmations and taking turns with the agent.

The \textbf{modality of user input} is another important factor to consider. Input modalities may include text, images, voice, and others, each of which can provide cues about the user's goals and the context of the interaction. For example, what a user hopes to achieve with an agent in a voice-controlled scenario, such as when experiencing situational impairments, can be different from their expectations when interacting with the same agent via keyboard input.

The \textbf{user profile} can be sent to the agent along with queries and serves as important context, particularly in cases of ambiguity. This profile may include factors such as the user’s activity history within the app, their familiarity with the UI, and their global preferences across multiple applications (e.g., dietary restrictions). An important design consideration is how to inform users about the profile data that may be used by the agent and how to give the user control over it.

\textbf{Ambiguity} in user queries refers to situations where the input is unclear due to user errors, missing parameters, or references to multiple possible elements on the UI. Different types of ambiguity may require different disambiguation strategies, involving distinct forms of interaction between the user and the agent.

Other \textbf{contextual factors} surrounding the user and their interaction with the agent should also be considered in the design of the user experience. These include the device type, the user’s goals, and their mental state. For example, a user may expect different interactions with the agent when they have a clear, specific goal in mind, versus when they are in an exploratory mode where they hope to discover options collaboratively with the agent.

It is also important to consider the \textbf{safety} of user queries. This includes developing a safety policy that defines what constitutes a safe or unsafe user query to the agent, and implementing proper guardrails to enforce safety policies, including mechanisms for detecting and defending against prompt injection and other forms of misuse.

\subsubsection{Explainability of Agent Activities}
Another important aspect of user experience for computer use agents is the explainability of agent's activities. This involves design decisions about what information to present to users and how to communicate various aspects of the agent’s activities. 

One key consideration is the \textbf{visibility of agent activities}---whether and where UI indicators reflect the agent’s actions. This includes how to present where the agent's actions occur within the UI. Another important aspect is retainability: is the history of the agent’s activities stored and accessible to the user? Can the user review past actions to understand or verify what the agent has done?

\textbf{Description of agent actions} involves communication of what the agent is doing. This includes descriptions of the actions being taken, and whether and how other tools or services are being invoked. Design choices around the level of detail and how much explanation to present to the user can vary based on factors such as available screen space on different devices.

\textbf{Transparency of agent reasoning} focuses on the design aspects around whether and how to present agent’s decision-making process to the user. This includes providing descriptions of the agent’s internal reasoning steps (often referred to as ``thoughts''), the knowledge sources consulted, and any uncertainty in the model’s understanding. 

\textbf{Preview of next steps} refers to informing the user about the immediate action the agent is about to take. Design considerations include how to present the thought process of the agent to reach to the decision for the next step and the various ways this preview can be presented to the user, such as through visual indicators on the UI element that is about to be clicked, or by presenting a textual description.

For agents that generate an overall plan based on the user’s query, it is important to design the \textbf{presentation of the plan} as part of the user experience. This includes how to initially present the full plan to the user and whether and how to communicate any changes due to replanning as the interaction progresses. 

\textbf{Communication of runtime status} refers to how the system informs the user about the agent’s current state, such as whether it is running, has completed a task, or is waiting for user input. It also includes providing confirmations of completed steps. A key design aspect is to consider the different scenarios where users actively monitor the agent’s progress (e.g., through a visible video or textual update) and where the user steps away or focuses on other tasks in parallel while the agent runs in the background.

\subsubsection{User Control}
This area of user control involves design choices around how users can intervene in the agent’s actions, both when the agent is performing correctly and when it makes a mistake. It also includes providing users with control options for high-risk actions, ensuring they can review, confirm, or cancel such actions before execution.

The types of \textbf{user intervention during agent execution} are perhaps the most immediate aspects of user control to consider in agent design. Our review of existing agents and insights from domain experts identifies several key intervention mechanisms, including basic controls such as stop and pause, the ability for the user to take over while the agent is executing, user demonstrations of example actions for the agent to generalize from, and the option to revert to a previous step and allow the agent to continue from there.

How users control the agent in \textbf{high impact scenarios} is another crucial aspect to consider when designing the user experience. Design considerations include providing indicators of potential risk or impact, effectively communicating the level of impact based on the context, and offering mechanisms for the user to grant or deny the agent permission to proceed. 

In addition to intervening while the agent is executing a task, another important consideration for agents that operate based on a generated plan is \textbf{user intervention on the plan}. This includes both intervention in the initial plan and during any subsequent replanning. Key design questions involve how users should be able to review and edit the initial plan, and how they should be informed about and involved in the replanning process.

In addition, user control in \textbf{agent error} situations must be carefully considered as part of the user experience. Errors can take different forms, including when the agent fails to follow the user's command or when it follows the command but the task fails due to interface limitations or other constraints. Important design questions include how to improve error discoverability and how to communicate errors effectively to the user.

\subsubsection{Mental Model}
Last but not least, an essential part of designing the user experience for an agent is helping users develop appropriate mental models and expectations about their interaction with the agent. We synthesize these considerations about mental model into the following key aspects:

How to inform the user about the \textbf{capabilities of the agent} is a key aspect of fostering appropriate mental models about the agent. This includes communicating the types of tasks the agent can perform, such as information retrieval, navigation and tutorials, configuration, executing transactions, or facilitating communication; explaining what kinds of user queries the agent can understand, how users can control the agent, and how the agent supports different types of user goals (e.g., executive versus exploratory tasks).

It is also important for user experience design to take into account the \textbf{UI context} itself, specifically, the user’s understanding of the interface, aspects of the UI that pose constrains on the agents' activities, and indicators on when the user interacts with the UI versus when the agent interacts with the UI. 

The \textbf{scope of the agent} refers to the boundaries of which part of the UI the agent can access and control. This can include the scope within the current application, across the device, and potentially beyond it. It also covers the external tools, services, and resources available to the agent outside the immediate application context. The design of the user experience should consider how to communicate this scope to users and enable the user to control the scope of the agent.

Another key consideration in design involves helping the user understand the \textbf{risks} that the agent can bring. This can involve considerations around what personal data and privacy-related information is accessible to the agent, and how users can control that access. This also includes identifying the types of risks the agent may pose to the user, and make the user be aware of and know how to employ the mitigation strategies to reduce those risks.

\subsection{Synthesis}
In this phase, we constructed the taxonomy describing the UX design space of computer use agents based on existing computer use agents and practitioner feedback. The four major categories, which define the general design areas of the taxonomy, echo usability considerations from early work on interface agents \cite{horvitz1999principle, lieberman1997autonomous, shneiderman1997direct} as well as more recent design guidelines for human–AI interaction \cite{amershi2019guidelines}. The 21 subcategories represent more specific aspects to consider within each design area and are adapted to computer use agents as an emerging form of human–AI interaction. Some areas captured by the taxonomy have also been identified as important for AI agentic systems beyond computer use agents, such as design dimensions related to safety, high-impact scenarios, and risk \cite{zhang2025from}; considerations around presenting plans and enabling user interaction with them \cite{feng2025cocoacoplanningcoexecutionai}; and the communication of agent actions and reasoning \cite{bansal2024challengeshumanagentcommunication}. Other dimensions are more specific to user interaction with computer use agents. Whether specific to computer use agents or applicable to general agentic systems, we expect the identified design dimensions to provide a shared vocabulary and useful directions to help designers reason about factors related to the user experience of computer use agents.
 
In this taxonomy, we do not aim to prescribe whether a particular design is inherently good or bad. Rather, we acknowledge that some aspects of the design space may be essential in some use cases while offering little or no benefit in others. %
Our goal is to provide a structured foundation that future work can build upon as the technology of computer use agents and its applications mature. In the next phase, we validate the taxonomy with empirical insights from a Wizard-of-Oz study.

\begin{figure*}[th]
    \centering
    \begin{subfigure}{0.265\linewidth}
        \centering
        \includegraphics[width=\linewidth]{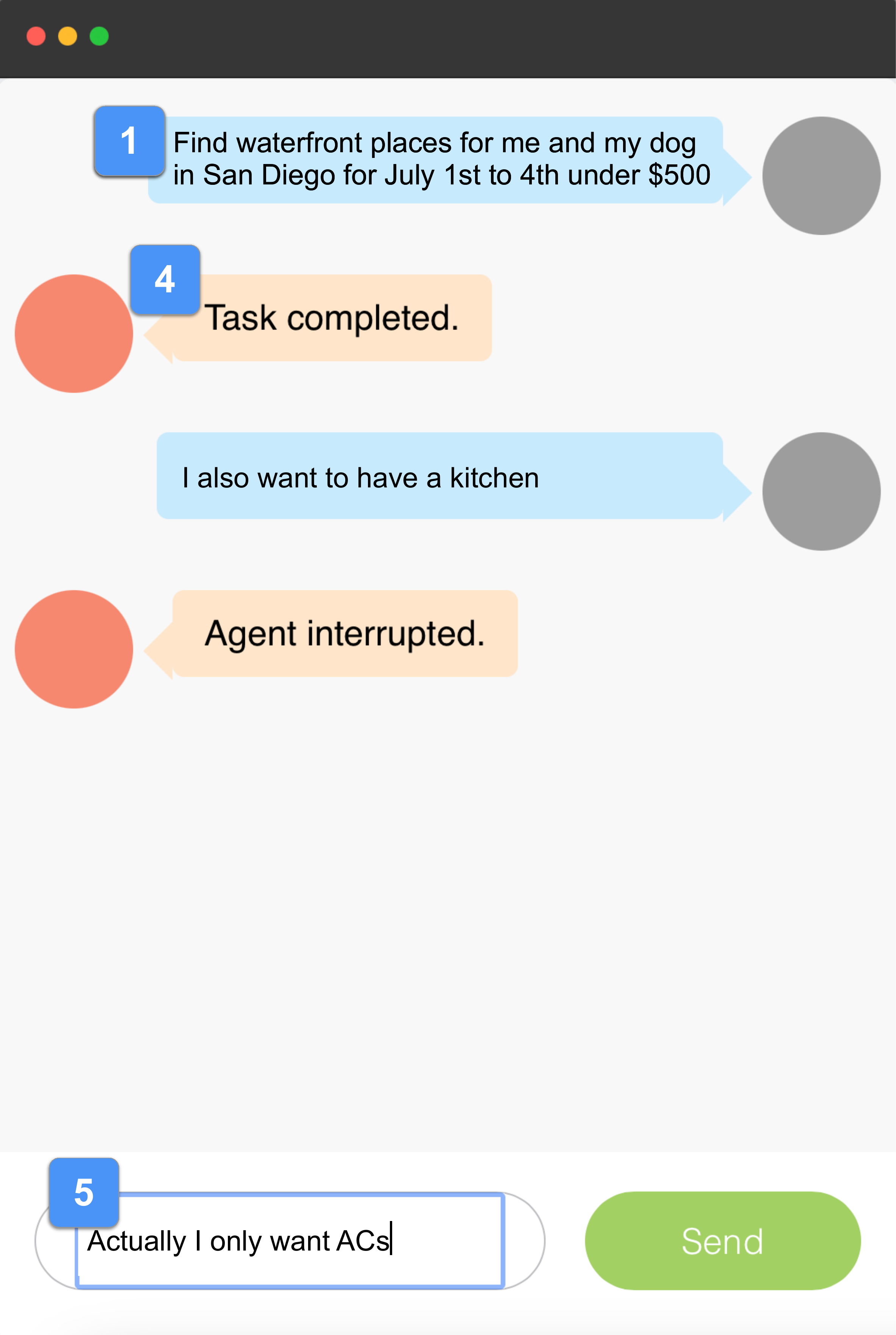}
        \caption{User chat interface}
        \label{fig:woz_userview}
    \end{subfigure}
    \hfill
    \begin{subfigure}{0.70\linewidth}
        \centering
        \includegraphics[width=\linewidth]{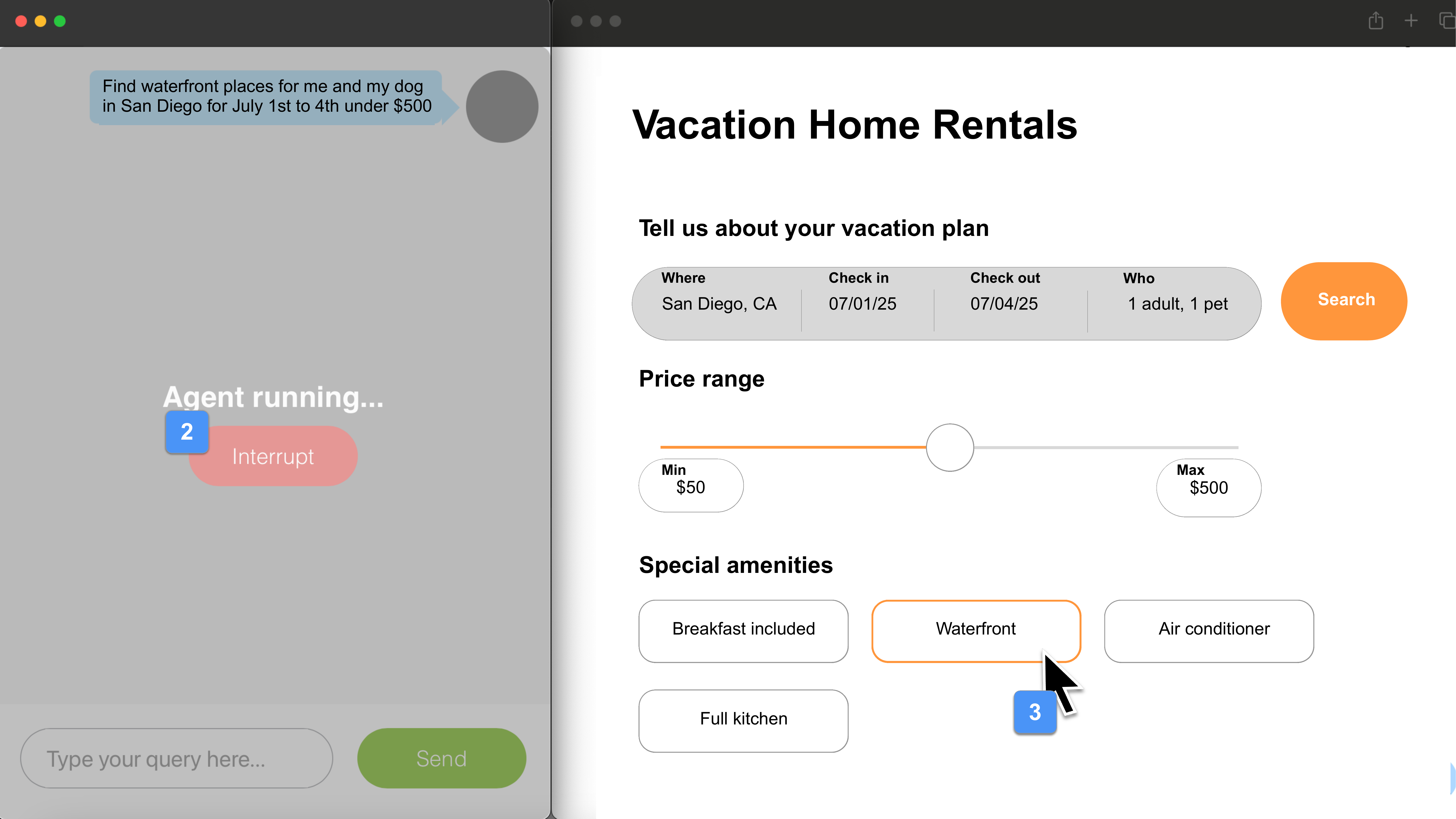}
        \caption{Interface for the Wizard-of-Oz agent execution}
        \label{fig:woz_modview}
    \end{subfigure}
    \caption{The Wizard-of-Oz study setup on a mockup example website.}
    \Description{Example illustration of the interfaces used in the Wizard-of-Oz study. The figure on the left is the User chat interface, which illustrates user and agent conversations (marked with number 1 and 4) and a text entry box (marked with number 5) for the user to enter their query. The figure on the right is the interface for the Wizard-of-Oz agent execution. On the left half of that figure is the chat interface, with a gray overlay indicating that the agent is running, and a interrupt button (marked with number 2). On the right half of that figure is a Vacation Home Rental web page, which contains sections of vacation dates, price range, special amenities, and a search button. A mouse controlled by the ``agent'' hovers on the page (marked with number 3).}
    \label{fig:woz_views}
\end{figure*}

\section{Phase 2: Evaluating the taxonomy and understanding user experience through a Wizard-of-Oz study}
While Phase 1 focused on building the taxonomy from existing computer use agents and practitioner feedback, 
in Phase 2 we conducted a Wizard-of-Oz study with 20 participants, in which the participants interacted with a ``computer use agent'' played by a researcher. Phase 2 had two goals: 1) To validate the coverage of the taxonomy; 2) To understand end users’ experiences and needs when interacting with computer use agents during normal execution, as well as in scenarios involving errors and risks.

\subsection{Method}
\subsubsection{Participants}
We recruited 20 participants from a large technology company through internal messaging channels that reached a wide range of teams, roles, and experience levels. To ensure a basic understanding of computer use agents, we used a screening survey to select individuals who had prior exposure, whether by watching demos, reading about existing agents, or using them directly. Our participants represented diverse job titles, including engineers, designers, and product managers. Participant profiles are provided in Table \ref{tab:woz_participants} in the Appendix. Our recruitment and data collection methods were reviewed and approved by a human research ethics committee and legal counsel at our company. Each participant received a \$24 gift card as compensation.

\subsubsection{Wizard-of-Oz study setup}
We conducted the study using a Wizard-of-Oz method \cite{dahlback1993wizard}. The researcher (the ``wizard'') and the participant were located in separate rooms and communicated via a video conference call. Participants were provided with a mock user chat interface (Figure \ref{fig:woz_userview}) through which they could interact with an ``agent'' played by the researcher. Meanwhile, the participant were also presented with the agent's execution interface, where the researcher acted as the agent and interacted with the UI on screen based on the participant's command (Figure \ref{fig:woz_modview}).

On the user chat interface, participants could enter textual queries in natural language (\nicenumber{5}), which then appeared in the chat thread (\nicenumber{1}). Then, the ``agent'' began execution, where the researcher controlled the mouse and keyboard on their end to simulate the agent’s actions on the web page (\nicenumber{3}). When the researcher completed the task, they entered a shortcut key that posted a ``task completed'' message in the chat thread (\nicenumber{4}). During execution, participants could use an interrupt button (\nicenumber{2}) to stop the agent, %
and a message ``agent interrupted'' would appear in the chat.

\subsubsection{Study procedure}
We began each study session by asking participants general questions to understand their experiences with computer use agents
and any challenges and concerns they had encountered.

Next, we introduced participants to a demo of our Wizard-of-Oz agent on a timer website. We explained that we have developed a basic browser-based agent that can take user commands (such as ``set a 15-minute timer'') and execute tasks through interacting with the UI on the webpage. %
After participants became familiar with the agent through the demo, we asked them to complete a series of tasks on a website using the agent. We encouraged them to think aloud while interacting with the system and to focus on the interaction aspects of the agent rather than on the underlying model’s performance.

Each participant completed tasks within one of two study scenarios: booking accommodations on a vacation rental website (\textit{Vacation Rental}) or purchasing items on an e-commerce website (\textit{Online Shopping}). Participants P1–P10 were assigned to the vacation rental scenario, while P11–P19 worked on the online shopping scenario. These scenarios were selected to represent typical use cases of computer use agents. Within each scenario, participants were prompted to complete six tasks using the agent (see Table \ref{tab:woz_tasks}). The tasks were designed to reflect a range of conditions identified in Section \ref{sec:taxonomy}: cases where the agent successfully completed the task, tasks with ambiguity, tasks involving agent errors, and tasks that carried potential risks or impacts for the user.
For agent errors, we included both abnormal runtime (e.g., the agent becoming stuck in a navigation loop) and normal runtime with mistakes (e.g., making a selection different from the user’s instruction). For impact-related tasks, we considered both monetary impacts (e.g., during financial transactions) and social impacts (e.g., when communication with others was involved), following \citet{zhang2025from}. The task set also covers a diverse range of user intentions outlined in the taxonomy from \citet{zhang2025from}, including information retrieval, navigation, configuration, executing transactions, and communication. Details of the specific tasks and corresponding agent actions are provided in Tables \ref{tab:woz_tasks_airbnb} and \ref{tab:woz_tasks_apple} in the Appendix.
At the end of each task, participants were asked to reflect on their experience interacting with the agent. 

At the end of the session, participants were asked to reflect on their overall user experience with the agent and to suggest any additional features they would like to see. Finally, we debriefed participants about the Wizard-of-Oz setup of the agent.

\begin{table*}[h]
\begin{tabular}{lll}
\toprule
\textbf{Task condition} & \textbf{Example scenario} & \textbf{User intention} \\ \midrule
Ambiguous task & \begin{tabular}[c]{@{}l@{}}User wants to explore an item on the page, \\ but there are multiple items of the same name\end{tabular} & \begin{tabular}[c]{@{}l@{}}Information \\ retrieval\end{tabular} \\ \midrule
{\color[HTML]{000000} Task correctly completed} & User wants to navigate to a specific product & Navigation \\ \midrule
{\color[HTML]{000000} Task running abnormally} & \begin{tabular}[c]{@{}l@{}}User wants to select a set of specs, \\ but the agent gets stuck in a navigation loop \\ as the specific option is unavailable\end{tabular} & Configuration \\ \midrule
{\color[HTML]{000000} Task running normally, completed incorrectly} & \begin{tabular}[c]{@{}l@{}}User wants to select a set of specs, \\ but the agent selects the wrong option\end{tabular} & Configuration \\ \midrule
Task with monetary risk & User wants to make a purchase & \begin{tabular}[c]{@{}l@{}}Executing \\ transactions\end{tabular} \\ \midrule
Task with social risk & User wants to communicate with someone & Communication \\
\bottomrule
\end{tabular}
\caption{Wizard-of-Oz study task table. }
\label{tab:woz_tasks}
\end{table*}

\subsubsection{Data collection and analysis}
We collected and transcribed video recordings of the study sessions as well as participants’ chat histories with the agent. To analyze the data, we followed a provisional coding qualitative analysis approach outlined by \citet{saldana2015coding}. The first author of this paper coded all transcripts using the subcategories in the taxonomy introduced in Section \ref{sec:taxonomy} and looked for any new subcategories. In addition, within each subcategory, the first author conducted open coding to iteratively develop themes that extended the example features in the taxonomy and conveyed why users considered particular aspects of user experience important (or unimportant). 
This process resulted in a total of $612$ coded quotes from all participants. 

In order to validate the coverage of our taxonomy, the second author of the paper coded a sample of $103$ quotes ($16.8$\% of the total dataset). These quotes were randomly sampled from the dataset, with five selected from each subcategory (except the \textit{Safety}, where all three quotes were selected). The second author first independently reviewed the subcategories assigned to the quotes by the first author and marked a set of $19$ quotes (18.4\%) that they would code differently than the original coding. The two authors then engaged in discussions to resolve disagreements and iterate on the coding, resolving $18$ of the $19$ quotes whose codes that the authors disagree on. 
Only $1$ quote (0.97\%) were decided as misclassified in the original coding and was corrected. Overall, this process validated the coverage of the subcategories in the taxonomy using the Phase 2 study data.

\subsection{Findings}

From our Wizard-of-Oz study, we did not find any new subcategories in this specific setup of the study, and that all 21 taxonomy subcategories were validated as important aspects of UX considerations during users’ interactions with an agent. Participants also provided details on each subcategory and explained how their preferences and expectations might vary depending on their goals and scenarios. These insights further deepen our understanding of the design space, which we present below.

\subsubsection{User query}

\paragraph{Differences in user queries with exploratory and specific goals.}
Users have different expectations regarding the types of queries they give to an agent and the kind of interaction they expect in return. When a user has a specific goal in mind, they may issue very detailed queries with clearly imagined steps, expecting the agent to treat it as a single request and carry it out from start to finish. For example, P15 described a case in which they wanted to automate certain activities on a website for repeated use, where they would want the agent to 
 ``do this very specific thing in this very specific order in this very specific way'' and ``be very prescriptive and verbose about what it's doing and and how it's doing it.'' When the user has specific goals, they also expect little interaction with the agent in the process, imagine it to be ``executing the task and maybe come back to the user in the end.'' (P15) 

 In contrast, when the user’s goal is more exploratory, they may not know the exact query to give the agent. In such cases, for example, as P19 described in a trip-planning scenario where they wanted to explore an unfamiliar country, the user prefers the agent to take on a more ``collaborative role'' to not only execute the given queries but also proactively offer suggestions and help users navigate the webpage in an exploratory way:
``I'm collaborating with an agent to build this mind map... there may be many options so it would be fine for me for it to grow as I go.'' (P19)

\paragraph{Varying expectations in agent autonomy towards ambiguous UI queries.}
A user query on UI actions can be ambiguous, whether unintentionally or intentionally, as a user query can refer to multiple UI elements appearing on a page. 
Users hold different views on how and when the agent should make decisions in ambiguous situations.

Some prefer the agent to make assumptions automatically in ambiguous cases, especially when the potential resulting divergence from user's intention is not critical:
``I think in the previous one, it felt like the paths kind of diverged too much.
Like they were two completely different items or products and it assumed the one that I wanted to choose... [versus] here, the way that the paths diverge in in this context doesn't feel that impactful.'' (P17)
User profile and context can help the agent make decision for disambiguation, as P5 explained with an example:
``I think of the model similar to having a personal assistant... If I told my assistant that, hey, I would like you to book this in this city, and there are multiple options for the city. I would expect my assistant to know which city I am in.''
Others like P2 imagined the agent could look at the users' calendar to figure out unspecified vacation rental dates. %
The user's previous interaction with the agent could also inform the agent decisions:
``how it has handled it for you in the past could inform how it handles it in the future.'' (P3)
At the same time, participants highlighted that such integration of user and contextual information in the query would require explicitly user control and permission, such as 
``a place where it could be like, `Oh, I know this about you, would you like me to go expand on this query?' '' (P2)

On the other hand, some users do not want the agent to make automatic decisions in ambiguous cases, especially when they are in an exploratory mindset, as P13 commented when deciding which product to explore:
``what I certainly would not want it to do is just assume an answer to one of those points of ambiguity because maybe I'm being misled. Maybe I was looking for [product A], but now you've taken me to [product B] but I missed the fact that there even was an [product A] and now I'm confused.''
Some participants, like P19, believe the agent should ``highlight areas where there is a decision point'' for the user to step in.
In those cases, users prefer the agent to proactively ask for clarification before making an action at the ambiguous step, such as
``...to stop and confirm, like `I'm on this page, I see multiple options, what would you like to choose?''' (P18)
Clarification can also happen when the user is making a query, where the system could
``process the prompt before the agent starts navigating, [to say to the user] that there could be multiple definitions for some of this stuff and then ask the user to clarify it before the navigation happens.'' (P17)
Alternatively, clarification can occur while the agent is presenting a plan to the user (P1).

\paragraph{Levels of expression serving user goals.}
In general, users prefer to command the agent with intentions rather than specific gestures, as it feels a ``more human way'' and ``humans more like[ly] [to be] aligned towards thinking in terms of tasks rather than steps.'' (P19)
However, in certain cases, users want to command the agent using specific gestures and have it follow exact steps. This is particularly important in educational tasks:
``[If] I want to explain some new UI to someone... I would say you click here, then you go here, you swipe until it goes like [this], so I would specify that.'' (P20)
Some users adopt a hybrid approach, starting with intent and moving to gestures only if necessary:
``I would always default to trying using [intent] first and if that didn't work out, in that case then I might try to get more defined.'' (P12)
In these cases, users emphasized the need for support in writing more specific, gesture-based prompts, such as features to help user
``break up the prompt into like more steps... It could just be like `when enter, creates a new step for the agent.' '' (P17)

\subsubsection{Explainability of agent activities}
\paragraph{Various attitudes towards visibility and monitoring of agent activities.}
Participants emphasized that an agent’s actions should always be visible on the screen. This visibility was seen as the key advantages of computer use agents compared to API-based agents: ``it could help the user feel more engaged and more confident that [what] the agent is doing is aligned into the task that the user intends.'' (P11)
Seeing how the agents operate on the UI can provide ``transparency'' (P19) in task execution, offering users ``a perception that human still have control'' (P19) over the tasks. In particular, seeing the cursor (i.e., which UI components the agent is operating on) allowed users to follow the agent’s actions: ``it just gives them all clear view about what's actually happening, where it's trying to click or figure out. It gives them all a clear picture of the action.'' (P4) 
Participants also emphasized the importance of having distinct visual indicators to differentiate between agent actions and user actions, such as a differently colored mouse for the agent.

At the same time, users generally agreed that they would be unlikely to monitor the agent in full detail in real world use, as doing so ``would not [be] saving a whole lot of time or energy'' (P3). Some participants mentioned that they would prefer to work on their own tasks while the agent runs in the background, particularly when the agent task is low risk, when the user has a specific goal in mind, or when the user feels confident in the agent’s performance.
To effectively monitor the agent when not actively watching it on the interface, participants suggested types of runtime information they would like to access, including an indication of progress (P1, P16, P20), the duration (P1), and status updates on whether the step was completed successfully (P7, P19).

\paragraph{Divergent needs in level of details in explanation}
Participants wanted to see descriptions of actions, reasoning, plans, and next steps at varying levels of detail depending on the scenario. For instance, when users prefer to step away and focus on their own tasks in parallel, they do not want to see detailed reports of the agent’s activities as they happen. Instead, they would rather receive a concise summary of the completed activities at the end for documentation purposes, as P20 noted: 
``I would like to have an option of don't report your intermediate status, just complete it [and] come back to me.''

Other participants suggested that an agent could share details about its activities during runtime but in a ``progressive disclosure'' manner (P15), which they explained as: ``by default it's telling very little or maybe only calling out the sort of top-level things that it's doing, and then maybe if the user wants to, they could drill in and see more.'' (P15)

There are scenarios where users want to see details about agent activities, especially when the agent makes decisions or inferences at points not explicitly specified in the user’s query. In these cases, users want to see the agent’s reasoning so they can understand how decisions are being made and remain in the loop: ``I don't wanna see things like there's only one button and I'm clicking on it. That doesn't give me any information. I think the information that I want is at any point that the decision needs to be made, a reasoning needs to be done, I want that surfaced back.'' (P19) Another important case is when a task carries potential risk or impact, and users prefer to be informed in real time to ``guarantee that it doesn’t do anything unexpected and [to] verify that it won’t do anything unexpected.'' (P17)

\paragraph{Scenarios where preview of plans and next steps are useful.}
There were also divergent opinions on whether and when users should see and be able to edit an overall plan for the agent’s activity before execution. Some argued that the plan should always be available for verification purpose as ``a way of checking that I was correctly understood.'' (P13)
This way, users could identify potential bugs or undesired steps in the plan, what P4 referred to as ``pre execution debugging'' and ``pre execution monitoring.'' For this purpose, participants emphasized that the plan should highlight high-impact or potentially risky steps, allowing users to confirm them in advance to mitigate risks:
``[Plan should show] the kind of risky step that is potentially destructive or can change my personal information or create an account or log in or make a payment. I would wanna have that called out and for me to approve.'' (P6)

Others felt that plans are only useful in certain cases. They argued that plans are most important when the agent is performing a task it has not done before, when the user is unfamiliar with the agent’s performance, or when the task involves certain risks. For example, both P7 and P8 emphasized the value of a plan for long tasks that users do not wish to monitor continuously:
``the only reason I want to do this (see the plan) is setting it (the agent) off on a lengthy task and then coming back to me with the results so I can go and do something else instead, so if I can quickly glance at its plan and say, ok, it's gonna open the website, it's gonna search this thing, it's gonna do that.'' (P8)
P15 further pointed out that plans are only useful in complex cases involving multiple steps and potentially different applications, such as ``booking flights and hotels and all these things across different tabs and different sites and domain[s].''
Along the same lines, participants like P4 noted that not all details need to be included in the plan and users should not be prompted to confirm every step, but they should retain the ability to deny actions if needed.

Beyond overall plans, participants emphasized the importance of informing users whenever replanning occurs during runtime, as P18 explained:
``having real time updates might be helpful cause you've already agreed on a plan and that's not actually happening, and maybe that's where the reporting is really important if you're doing something like buying something and the agent made random decisions.'' (P18)
The specific deviations from the original plan should be highlighted and explained, either in context or as a summary once the task is complete, as P2 suggested:
``highlight the fact that [it] has done it like in context, in the moment, or like at the very end to be like, hey, I got you to this point, but FYI I changed this thing.'' (P2)

\subsubsection{User control}

\paragraph{Varying levels of user engagement during agent runtime.}
We observed several ways in which users preferred to engage with the agent while it was running.
Many participants used the interrupt button on our Wizard-of-Oz interface, particularly when the agent made mistakes or became stuck in a loop, and they commonly wanted to follow up with a new command. For instance, when the agent could not find the correct color yellow of a product to match a user’s query and became stuck, P18 explained that they wanted to ``stopped it and say, hey, just ignore [it] is not yellow, that's fine. Move on.''
Some participants also imagined scenarios where they could temporarily take over control from the agent and then hand it back:
``once you did the thing (taking over the task)... and then you click the `I'm done button' and then it tells [the agent] basically what it [can] figure out from or what it [can] infer from you. '' (P2)
In those cases, the transition of control between the user and the agent needs to be obvious, such as what P3 suggested: ``having significant UI moments where your screen is kind of dimmed or there's like some layering or something to kind of indicate something's happening on your behalf or you've taken it back.'' (P3)

However, some participants noted that user intervention mechanisms would be ineffective if users were not paying attention to the agent’s actions during runtime. As P17 explained, the interrupt button is ``definitely not enough if we assume that the user is not staring at this running the whole time or that the navigation is not really visible like as much.'' %
An interesting idea raised by participants was that the agent could potentially detect whether a user is paying attention and adjust the level of control or confirmation requests accordingly. Especially in low-risk steps, participants imagined that the agent might engage more actively with the user through dialogue if the user is present and wishes to be involved; Conversely, if the user is not present, the agent could take initiative, make decisions on their behalf, and later report back those decisions (P15)
In contrast, when tasks involve risk, such as social risks in communicating with people, participants stressed that the agent should always hand control back to the user, even if the user is not actively present, as well as reminding the user about the risks.  We discuss this theme more in the next section.

\paragraph{Involving user confirmation in high impact scenarios.}
User control of the agent is particularly important in situations where risks are involved. Participants consistently emphasized the need for explicit user permission on irreversible tasks, as P2 shared,
``I don't want you to do anything that's irreversible, like I don't want you to hit the purchase buy button for something that's not cancellable, and to always ask me before you do that.'' (P2) 
In another example where the agent was making a voice call to a customer representative on the user’s behalf, participants also emphasized the need for confirmation before the communication was initiated:
``that was a bit scary like it [is] going to be a voice call... I think what I would expect it here is surfacing back information like: `there is an option to get help by chatting live with the specialist, is this now a good time to connect you?' '' (P19)

Meanwhile, participants also envisioned different tiers of impacts that should correspond to different types of agent interaction. 
Tolerability varied depending on the level of risk or potential impact, for example, the amount of money involved in a payment: 
``having the agent execute payments for you, for smaller items like pizzas or food deliveries and stuff, I don't think people will hesitate once they've had the confidence of it working correctly the first few times. For large items, I can understand people might be more hesitant... dependent right on how much money they're willing to risk.'' (P11)
In other words, participants preferred to engage with the agent only at points that could have meaningful consequences, as P5 explained,
``I don't want it to tell me anything until it gets to the point at which I have to click finish... I'd want it [to] finish doing the login, adding the payment method, all of that stuff, but just prior to finish like the final step.'' (P5)

\paragraph{Different ways to discover and recover from agent errors.}
On one hand, users acknowledged that the agent can make errors; on the other hand, such errors can sometimes undermine trust, particularly when they involve hallucinations (P5).
Users' tolerance for errors depended on the context, and they were more forgiving when the task was exploratory in nature, as P9 noted,
``typically if it's more specific, you would expect it to succeed right or to be able to do what you're asking.
And when it's more general, it's probably not surprisingly more prone to making an error or maybe not doing what you wanted it to do.''

For error discovery, several participants noted that if the user was not paying full attention during runtime, errors could easily go unnoticed.
Therefore, when users were not actively monitoring the agent, they wanted to be notified about any problems. This would require the agent to have the capability to self-evaluate its action and detect errors. 
For example, before an error occurred, the agent could proactively seek clarification to avoid getting stuck, as P17 proposed,
``maybe more important [for the agent] to come back to the user and be like, I don't see what you said. What do you mean and then maybe some options---please select your choice.'' (P17)
Participants also suggested adding UI indicators on the screen to highlight where the error occurred, as well as providing textual descriptions of the error (P9). 

For error recovery, in addition to interruption and manual correction, users also envisioned mechanisms that would allow them to revert the agent to a prior step. 
For instance, P17 highlighted a scenario in which the agent might click on the wrong webpage and imagined ability to return to the previous page and correct the error without losing progress:
``hopefully it would save internet history [and] I can go back for undo[ing] last action.'' (P17)
To improve the experience of reverting actions, participants suggested adding reversibility action explanations that highlight alternative paths the agent did not take, as P10 explained:
``there was another path that we could have gone down, but this is the one we did go down so that I know how far to back up... I don't wanna start over to have to repeat my entire process... I would want to be able to say wait, wait, wait, go back up a level.''
Others recommended incorporating visual representations of the agent’s decision steps to better support reverting actions. For example, P20 suggested
``a visual representation of the tree and all of the options and where on which branches that the agent went,'' so that they can tell it to ``actually start from here, go back and choose a different branch.'' (P20)

\subsubsection{User mental model}
\paragraph{Agent capability contributing to user trust.}
Participants highlighted the importance of helping users build accurate mental models of the agent’s capabilities. For example, P1 suggested ``some education screen in the beginning... kind of like showing people how to use it, [and] what level of specific detail is required.'' Observing the agent’s actions in real time was seen as a way for users to better understand what the agent can and cannot do, as P2 explained: ``if I saw it scrolling through this entire list and I would know, oh, it's only looked at these six [rental options]... being able to watch it would tell me a lot about like how it has done.'' 

Understanding the capability of the agent contributes to building appropriate trust with the agent. To this goal, it is also important to make it clear what aspects of the agent users can control. 
Trust was also linked to how the agent communicates its limitations, such as P19’s suggestion that ``they (agents) would basically say that I am not able to do that or I am not allowed to do that.'' Expressing uncertainty and asking for clarification was further seen as a key feature that can give users confidence in the agent, especially in risky cases. For instance, P14 explained: ``if it asks for clarification in cases where it's not sure, like, oh, hey I'm about to spend money, are you sure you want to do this? ... I think that helps a lot for it expresses it has some logic around `I will only use this information on this exact website for this'.''

\paragraph{UI context shaping user interaction with agent.}
Participants broadly agreed that their own familiarity with the UI of an application shapes their expectations for how to interact with an agent on it. When the interface is well-understood, users are comfortable delegating execution to the agent. But when they are unfamiliar with the page, even in a scenario with low risks, they want the agent to show more transparency, such as showing intermediate steps, explaining actions, and pausing for confirmations. P20 explained this nuance in an example: ``let's say that you buy a [bus service] card and you need to register it. And I've never done that... %
Given that I've never done it, and this is not a system that I'm familiar with, there might be some hiccups, so I would want to make sure that everything is right. So this is not a high stakes situation, but it's something that requires both this accurate information and the fact that I've never done this myself, I would want it to go step by step.''

\paragraph{Engaging user with agent scope to enhance privacy and mitigate risks.}
Participants emphasized the importance of clear indicators showing which parts of an application the agent has access to. 
To this end, user needs explicitly define the scope of the agent on data access, with P1 explaining:
``I also would want users consent on knowing what applications or what kinds of data the system should be able to see... like you know when you download an app, you have to basically say, does it have access to your folders, does it have access to your contacts. (P1) 
Similarly, P13 stressed the need for clear limitations on what the agent should not access:
``maybe there are some apps that I absolutely do not want it to ever use. Maybe my health information---like I don't ever want you to open health app.'' (P13)

Participants imagined control of the agent’s scope during the initial setup stage when granting overall permission, as well as dynamically at runtime when adjustments may be needed. For instance, P13 highlighted the need for informed control of the scope during setup, such as ``global levels of permission that the user provides... explaining when and why and how that that particular application would be used and allowing the user to either opt into that capability and therefore opt into that permission.'' 
Beyond initial setup, participants also wanted the ability to manage scope during runtime. 
For example, P12 pointed out that when the agent is executing a task that requires expanding its access to from browser to some other application, or an OS level, it should require explicit approval: ``[if] it's not contained to the browser session, it basically got free run of your system. I think there need to be guardrails on what it's capable of doing. Like, I don't know if I would want it to necessarily enable my microphone without my additional explicit authorization.'' 
Finally, participants highlighted the value of runtime indicators that made the scope visible, such as P16’s suggestion on ``some helpful visual indication about the windows where it can live, [the window] that I don't want it to touch.''

\section{Discussion}

While much of the current research has focused on model design and evaluation benchmarks, our paper is the first to examine the space of user experience with computer use agents in depth. We contribute a taxonomy that identifies four major areas shaping the user experience of computer use agents: user query, explainability of agent activities, user control, and user mental models \& expectations. Within these areas, we propose 21 dimensions (i.e., subcategories) that developers can consider when designing user experiences for computer use agents. %
We validated this taxonomy through a Wizard-of-Oz study and further described the divergent end-user scenarios and needs regarding each design areas.  %

Notably, the design dimensions listed in the taxonomy are explicitly non-opinionated. We do not intend for this taxonomy to provide guidance on whether a particular design is good or bad, nor on what should be prioritized over other considerations. We believe the field of computer use agents is too nascent, rapidly evolving, and insufficiently mature to confidently offer prescriptive guidelines at this time. As shown in the Phase 2 study, aspects of design captured by the taxonomy may be essential for one use case while offering little or no benefit in another, and thus require additional context-dependent investigation before any definitive recommendations can be made. 
Furthermore, we do not claim that this taxonomy is completely exhaustive. As discussed in Section \ref{sec:limitation}, there are many directions along which it may be extended. We invite future work to identify use cases of computer use agents that are not included in this taxonomy, and to contribute new dimensions that extend and refine it.

In the following sections, we discuss some key takeaways from our studies by connecting our findings with design considerations about pre-LLM era interface agents and general human-AI interaction systems. We invite developers to use our taxonomy as a map and a list of vocabulary for what to consider about user experience in the design of computer use agents, and to reflect on, experiment with, and account for the potential trade-offs regarding the diverse user needs and scenarios.

\paragraph{Handling ambiguity about UI elements in user queries:}
As \citet{horvitz1999principle} suggested in the critical factors of mixed initiative interface agents, agents should handle ambiguity by ``considering uncertainty about a user’s goals'', ``inferring ideal action in light of costs, benefits, and uncertainties'', and ``employing dialog to resolve key uncertainties.'' 
Our findings from both the taxonomy development and the Wizard-of-Oz study echo this early vision, and we extend it by providing details in the context of (M)LLM-based computer use agents. Ambiguous queries can occur in cases where a single query may refer to interactions with multiple UI elements on the screen, and such ambiguity may stem from different levels of user expression (e.g., expressing intent rather than gestures) or from the user’s goals (e.g., during exploratory UI tasks).  
When designing computer use agents, it is critical to consider how and when the agent should make decisions on which UI element to interact with when the user query is ambiguous, and how the agent should adapt to different contexts and user goals. For example, users may prefer the agent to make reasonable assumptions automatically when the risk of taking a diverting UI path is low, such as navigating to a webpage section through search instead of through navigation menu. However, in high-stakes or impactful situations, such as those involving payments, users do not want the agent to act automatically and instead prefer to be informed of possible options and have the agent request clarification before proceeding with an action.

\paragraph{Tensions around user engagement with computer use agents:}

Since the early days of interface agents, researchers have debated how and when users should be involved in the agent execution process. For example, Horvitz \cite{horvitz1999principle} argued that interface agents should ``consider the status of a user’s attention in the timing of services'' and ``provide mechanisms for efficient agent–user collaboration to refine results.'' Similarly, Shneiderman \cite{shneiderman1997direct} discussed the advantages and disadvantages of enabling users to directly manipulate agent activities.
Our work revisits these discussions in the context of (M)LLM-based computer use agents and uncovers new tensions around user engagement. Specifically, we identify two forms of engagement:
reviewing agent activities (e.g., explanations of the agent’s actions, reasoning, and plans) and 
controlling agent activities (e.g., stopping or reverting actions, confirming and approving steps). These forms of engagement echo the varying levels of user involvement in LLM-based systems identified by \citet{feng2025levels}---observer, approver, consultant (who provides feedback), collaborator (who co-plans and executes with the agent), and operator (who directs and makes all decisions). However, our study of computer use agents suggests that users are not satisfied with remaining in a single role; instead, they need to dynamically switch between reviewing and controlling the agent depending on context and tasks.
While our findings align with recent studies showing that users value mechanisms for reviewing and controlling agent plans \cite{he2025plan-Then-Execute, feng2025cocoacoplanningcoexecutionai}, how and when such mechanisms should be offered needs careful design consideration. Some users prefer not to monitor these actions during execution and are concerned that the multimodal agent runtime information (e.g., textual descriptions, screen recordings, or other content) could lead to information overload. At the same time, without active engagement, users might miss critical steps or errors made by the agent.
These findings echo \citet{bansal2024challengeshumanagentcommunication}, which lists ``choosing appropriate levels of detail'' as a central challenge in human–agent communication. 
While our study does not offer immediate solutions, the development of computer use agents should take these tensions into account and analyze the trade-offs in the level and type of information presented to user about the agent. %

\paragraph{Fostering appropriate trust in automatic UI actions:}
Trust has long been recognized as a critical dimension of human–AI collaboration \cite{amershi2019guidelines, bansal2024challengeshumanagentcommunication}. It is particularly important in the context of computer use agents. Agents performing UI actions can introduce varying levels of risks, as some actions are irreversible and may result in real-world consequences such as monetary loss or damage to one’s social image \cite{zhang2025from}. 
Our taxonomy and Wizard-of-Oz study echo the urgency of designing for appropriate trust in computer use agents and highlight several important factors to foster. First, it is crucial to inform user about the scope of the UI that the agent can access and how user can control it. Second, it is vital to engage users with the agent before the execution of impactful steps (e.g., making purchases or interacting with real people). Finally, although the agent may have the ability to operate without continuous user attention and automate UI actions, our findings echo recent developments in explainable AI, suggesting that agent designs could intentionally embrace ``seamfulness'' \cite{ehsan2024seamful}. Such designs should prioritize user understanding and preserve users’ agency to intervene, particularly in situations involving ambiguity and uncertainty. Applied to the user experience of UI agents, this could involve explicitly indicating when the agent lacks sufficient knowledge rather than making automatic assumptions, thereby allowing users to better calibrate their expectations and level of real-time engagement.
Future work could further explore these directions to help users calibrate trust in computer use agents.

\subsection{Limitations and Future Work} 
\label{sec:limitation}
This work is qualitative and exploratory by its nature, and therefore has several limitations. Although the taxonomy developed in Phase 1 is intended to be platform-agnostic, Phase 2 focused exclusively on Wizard-of-Oz scenarios for browser-based tasks, which limits insights about its applicability to other platforms. In both phases, all participants were recruited internally from our company, which may introduce demographic and professional bias toward tech-savvy, well-educated, English-speaking populations, thereby limiting external generalizability. As a result, some important applications of computer use agents, such as accessibility, are not examined in this study. Furthermore, this internal recruitment method, driven by company policy, may also lead to bias and responses influenced by social desirability. In Phase 2, because the agent is simulated, certain user experience considerations related to real-world model performance, latency, or multimodal perception may not fully manifest. Recognizing these limitations, this formative work opens up many directions for future research, which we discuss below. 

While our Wizard-of-Oz study revealed how end users of computer use agents perceive the design dimensions identified in the taxonomy, we intentionally refrained from making claims about how the experience of computer use agents should be designed across the design space. Instead, our study uncovered tensions that even for the same design dimension users expressed different needs depending on factors such as context and goals, such as the various levels of user engagement during runtime, the different expectations regarding agent autonomy towards ambiguous UI queries, and the divergent needs in level of details in explanation of agent activities. Future research could treat these findings as directions for hypotheses and employ controlled experiments or other quantitative methods to test support for specific design directions.

Our taxonomy provides an initial starting point for designers and developers to consider the user experience of computer use agents, but variations may emerge depending on the specific characteristics of the agent. For instance, although our taxonomy aims to outline a platform-agnostic design space, different platforms, as highlighted by the \textit{input modality} subcategory, may result in distinct design considerations. We encourage future researchers to adapt and extend the taxonomy with additional validation on mobile,
desktop, or OS-level agents. In parallel, research on tooling for computer use agent development could investigate ways to recommend key design dimensions that designers and developers should prioritize for specific use cases, such as by providing scaffolds for each relevant design dimension~\cite{liang2025agentbuilder}. In addition, further studies with more diverse populations (e.g., non-technical individuals, older adults, non-English-speaking individuals, and people with disabilities) could help assess the generalizability of our findings and inform future extensions of the taxonomy.

An important dimension not captured in our taxonomy is accessibility. Although it was beyond the scope of this project, we recognize its significance, particularly given the great potential of using AI-powered UI control agents as accessibility features \cite{vu2023voicify}. Future research could extend the taxonomy to include accessibility, for example, by exploring how it applies to the needs of visually impaired users when interacting with agents.

\section{Conclusion}
In this two-phase formative study, we explore the design space of user experience for computer use agents. In Phase 1, we developed a taxonomy that maps the general design space of UX for such agents, drawing from an analysis of nine recently announced agents and interviews with eight practitioners in UX and AI. The taxonomy highlights key areas of design consideration, including: what to account for when supporting user queries; which aspects of the agent’s activities should be explained to the user; how to grant users appropriate control over different parts of the interaction; and what constitutes the user’s mental model of the agent.
In the Phase 2, we conducted a Wizard-of-Oz study with 20 participants, where a researcher acted as a web browser–based computer use agent to carry out tasks related to vacation rentals and online purchases based on user commands, probing for user reactions and needs both during normal execution and in situations involving errors or risks. The findings validate the taxonomy from Phase 1 and enrich the design space with deeper insights into user experience.
Through this two-phase approach, we contribute both a taxonomy and empirical findings on UX for computer use agents. Future designers and developers can refer to this taxonomy to systematically account for different aspects of user experience in the design of computer use agents.
We also invite future work to identify additional use cases of computer use agents and to extend this taxonomy with new dimensions as the technology evolves.

\bibliographystyle{ACM-Reference-Format}
\bibliography{reference}

\newpage
\newpage
\onecolumn
\appendix

\section{Appendix}
\subsection{Phase 2 participant profile}
\begin{table}[h]
\begin{tabular}{lll}
\toprule
\textbf{Id} & \textbf{Scenario} & \textbf{Job role}   \\
\hline
P1          & Vacation rental   & Engineer        \\
P2          & Vacation rental   & Designer        \\
P3          & Vacation rental   & Designer        \\
P4          & Vacation rental   & Engineer        \\
P5          & Vacation rental   & Engineer        \\
P6          & Vacation rental   & Engineer        \\
P7          & Vacation rental   & Engineer        \\
P8          & Vacation rental   & Product manager \\
P9          & Vacation rental   & Engineer        \\
P10         & Vacation rental   & Engineer        \\
P11         & Online shopping     & Engineer        \\
P12         & Online shopping     & Product manager \\
P13         & Online shopping     & Designer        \\
P14         & Online shopping     & Engineer        \\
P15         & Online shopping     & Engineer        \\
P16         & Online shopping     & Product manager \\
P17         & Online shopping     & Engineer        \\
P18         & Online shopping     & Engineer        \\
P19         & Online shopping     & Engineer        \\
P20         & Online shopping     & Engineer       \\
\bottomrule
\end{tabular}
\caption{Wizard of Oz study participant profile}
\label{tab:woz_participants}
\end{table}

\newpage
\subsection{Phase 2 Wizard-of-Oz tasks}

\begin{table}[h]
\begin{tabular}{lll}
\toprule
\textbf{Task condition}                                                                 & \textbf{Task scenario}                                                                                                                                                      & \textbf{User intention}                                          \\ \midrule
Ambiguous task                                                                          & \begin{tabular}[c]{@{}l@{}}Find results for a trip to Springfield \\ (Multiple Springfield options)\end{tabular}                                                            & \begin{tabular}[c]{@{}l@{}}Information \\ retrieval\end{tabular} \\ \midrule
Task correctly completed                                                                & \begin{tabular}[c]{@{}l@{}}Find vacation home in San Diego for 2 adults and a dog \\ from 9/1/25 to 9/3/25\end{tabular}                                                     & Navigation                                                       \\ \midrule
Task running abnormally                                                                 & \begin{tabular}[c]{@{}l@{}}Only look for homes that are at waterfront under \$500 per night \\ (Agent cannot find waterfront options and gets stuck in a loop)\end{tabular} & Configuration                                                    \\ \midrule
\begin{tabular}[c]{@{}l@{}}Task running normally, \\ completed incorrectly\end{tabular} & \begin{tabular}[c]{@{}l@{}}Choose a home from 9/1/25 to 9/5/25 \\ (Wrong dates are selected)\end{tabular}                                                                   & Configuration                                                    \\ \midrule
Task with monetary risk                                                                 & Make a payment for the reservation                                                                                                                                          & Executing transactions                                           \\ \bottomrule
\end{tabular}
\caption{Tasks used in the \textit{Vacation Rental} scenario on the \url{www.airbnb.com} website.}
\label{tab:woz_tasks_airbnb}
\end{table}

\begin{table}[h]
\begin{tabular}{lll}
\toprule
\textbf{Task condition}                                                                 & \textbf{Task}                                                                                                                                                       & \textbf{Task type}                                               \\ \midrule
Ambiguous task                                                                          & Learn about Apple TV (Multiple Apple TV options)                                                                                                                    & \begin{tabular}[c]{@{}l@{}}Information \\ retrieval\end{tabular} \\ \midrule
Task correctly completed                                                                & \begin{tabular}[c]{@{}l@{}}You have a niece and you want to buy an iPad for them. \\ They likes the color yellow and the smallest storage is fine\end{tabular}      & Navigation                                                       \\ \midrule
Task running abnormally                                                                 & \begin{tabular}[c]{@{}l@{}}You want to compare it with iPad mini of the same color \\ (Agent cannot find color yellow option and gets stuck in a loop)\end{tabular} & Configuration                                                    \\ \midrule
\begin{tabular}[c]{@{}l@{}}Task running normally, \\ completed incorrectly\end{tabular} & \begin{tabular}[c]{@{}l@{}}Get the starlight color iPad mini with a warranty \\ (Wrong warranty option are selected)\end{tabular}                                   & Configuration                                                    \\ \midrule
Task with social risk                                                                   & \begin{tabular}[c]{@{}l@{}}You want to ask for more guidance about warranty \\ before making the purchase \\ (Agent opens customer service chat)\end{tabular}       & Communication                                                    \\ \bottomrule
\end{tabular}
\caption{Tasks used in the \textit{Online shopping} scenario on the \url{www.apple.com} website.}
\label{tab:woz_tasks_apple}
\end{table}

\end{document}